\documentclass[aps,prb,twocolumn,showpacs,superscriptaddress]{revtex4}
\usepackage{epsfig}
\usepackage{times}
\usepackage{amsmath}
\usepackage{url}
\usepackage{color}

\begin{document}

\title{Mapping the Arnold web with a GPU-supercomputer}

\author{A. Seibert}
\email[Electronic address (corresponding author): ]{armin.seibert@physik.uni-augsburg.de}
\author{S. Denisov} 
\author{A. V. Ponomarev}
\author{P. H\"{a}nggi}
\affiliation{Institute of Physics, University of Augsburg,
Universit\"{a}tstr.~1, D-86159 Augsburg, Germany}
\date{\today}
\begin{abstract}
The Arnold diffusion constitutes a dynamical phenomenon which may occur in the phase space of a non-integrable Hamiltonian system whenever the number of the system degrees of freedom  is $M \geq 3$. The diffusion is mediated by a web-like structure of resonance channels, which  penetrates the phase space and allows the system to explore the whole energy shell. The Arnold diffusion is a slow process; consequently the mapping of the web presents a very time-consuming task. We demonstrate that the exploration of the Arnold web  by use of  a graphic processing unit (GPU)-supercomputer can result in distinct speedups of two orders of magnitude as compared to standard CPU-based simulations.
\end{abstract}

\pacs{05.45.Ac, 01.65.+g}

\maketitle

\textbf{
Several  archetype results of dynamical chaos theory can  unambiguously be attributed to unforeseen outcomes of numerical experiments. Two such examples are the absence of thermalization in the Fermi-Pasta-Ulam (FPU) chain \cite{FPU} and the discovery of the Lorenz attractor \cite{lorenz}. In some cases the  use of numerical simulations presents the only method to obtain insight into the behavior and evolution of the system of interest. There are several branches of modern physics, which are computational by their own nature, with Computational Cosmology \cite{CC} being such a paradigm.
Nowadays, computational cosmologists perform simulations with more than $10^{10}$ particles\cite{nature}. The main driving force underpinning this advance is the high \textit{parallelization} of simulations that allows one to run an artificial Universe on thousands of processors simultaneously. On the other hand, one may  benefit as well from a high parallelization on the scale of much smaller systems. For example, the averaging over many realizations of a stochastic force or of quenched disorder to arrive at scaling exponents, the Monte-Carlo sampling etc., present  typical routines in many areas of computational physics. It is evident that by running $N$ different realizations on $N$ processors in parallel, one can speed up the statistical collection process by a factor of $N$. One of the possibilities is then to use a computational cluster by sending the  program to run on many CPUs simultaneously, collecting the output data and finally sample them.  This ``sending-collecting-sampling'' process is cumbersome work, which, however, could be avoided by using special designed scripts. The advent of Graphics Processing Units (GPUs) has brought such numerical experiments into a new level of performance \cite{gpu1}. Originally used as hardware chips,  designed as graphic data-pipelines, GPUs  were soon recognized as an additional beneficial tool:  researchers from such areas  as medical imaging, computational electromagnetics, and hydrodynamics, have successfully implemented them for data processing \cite{nvidia}. Nowadays computational physics is marked by an impressive boost of the  ``General Purpose Computing on GPU'' (GPC--GPU) ideology \cite{gpc}. With this work we attempt to demonstrate how the use of a GPU-supercomputer can provide  useful insight for complex problems of nonlinear dynamics such as  Arnold diffusion \cite{arnold}.}

\section{Introduction}

The Arnold diffusion is not that kind of phenomenon that involves the time evolution of billions of particles. In fact, it can appear in the phase space dynamics of a Hamiltonian system whenever the number of system degrees of freedom is $M \geq 3$. The Arnold diffusion is relevant in celestial mechanics and astronomy \cite{astro}, plasma dynamics in stellarators and tokamaks \cite{accel, accel2}.  It also influences the evolution of a Rydberg atom placed in crossed magnetic and electric fields \cite{rydberg},
or it might explain experimentally observed effects of emittance growth in the TeVatron colliders \cite{emmit}. Moreover, Arnold diffusion may be a mechanism of the ultraviolet cut-off in statistical mechanics \cite{stat}. Except some specially designed models, where the presence of the diffusion can be rigorously proved and the diffusion timescales can  theoretically be estimated, only little work is generally  analytically possible. Therefore, numerical experiments play an prominent role in the studies of Arnold diffusion \cite{numericA, numericAa, numericA1, numericB}. However, conclusive numerical output then requires extremely large time scales; -- because the actual rate of the process is not known {\it ab initio}. With this study we demonstrate that the presence of the Arnold diffusion in the dynamics of a  model Hamiltonian system can be visualized by scanning the system phase space with a giant ensemble of trajectories. This has been realized within the Compute Unified Device Architecture (CUDA) framework \cite{cuda_zone} with its benchmarks performed on a NVIDIA Tesla GPU. We detect the Arnold web \cite{web}, i.e. a chaotic network which can carry the system over the  energy shell, in the limit of extremely weak perturbations. We also resolve a rich fractal structure of the Arnold web in the regime of moderate non-integrability, when the clusters of high-order resonances start to contribute to  diffusion.

\section{Model for Arnold Diffusion}

One of the typical models to study the Arnold diffusion is a system of coupled standard maps, see, for example, Refs.\cite{smap}. Such systems are easy to handle because they only need to be iterated and do not demand sophisticated integration schemes. However, they represent a class of \textit{driven} Hamiltonian systems, and their exposition to a train of delta-kicks leads to an energy pumping and an unlimited diffusion in the momentum subspace. Therefore, time evolution of such systems is not restricted to a compact manifold in the corresponding phase space. 

Here we want to address the case of \textit{autonomous} Hamiltonian systems. Hamiltonians of such systems represent intergals of motion, and their time evolution is restricted to  manifolds of constant energy.

An autonomous Hamiltonian system with one degree of freedom, $M = 1$, is always integrable since it possesses the integral of motion,  $E = H(x,p)$. Hamiltonian systems with two degrees of freedom, $M=2$, evolve in a four-dimensional phase space, $\mathbf{\Omega} =\{(x,y,p_x,p_y)\}$, but the system evolution is confined to the energy shell of  energy $E=H(x(0),y(0),p_x(0),p_y(0))$,  determined by the initial conditions. Regular two-dimensional invariant manifolds, \textit{tori} \cite{web}, separate the system energy shell into different regions. Different regions exhibit different dynamics, chaotic or regular ones, but each region is perfectly isolated from the remaining ones. This is so because the two-dimensional tori   provide a complete partition of the three-dimensional energy shell. This topological argument does not work anymore in higher dimensions: the $M$-dimensional tori cannot partition the $(2M-1)$-dimensional shell  whenever the number of the degrees of freedom $M \geq 3$.
Therefore, there could be  trajectories that slip between regular tori and thus are able in exploring the whole energy shell.

To be more specific, we consider a three-dimensional system with a Hamiltonian:
\begin{equation}
H(\mathbf{P},\mathbf{X})=  \frac{\mathbf{P}^2}{2} + \varepsilon H_p(\mathbf{X}) = H_0(\mathbf{P}) + \varepsilon H_p(\mathbf{X}),
\label{eq1}
\end{equation}
where $\mathbf{P} = (p_x, p_y,p_z)$ and $\mathbf{X} = (x, y, z)$ denote the momentum and coordinate vectors.
For $\varepsilon = 0$ the system is completely integrable, and the vector $\mathbf{P}$ is constant along any trajectory of the system,
$\mathbf{X}(t) = \mathbf{X}(0) + \mathbf{P}t$. For a given energy $E$ the energy shell forms a sphere  in the momentum subspace, $\mathbf{S}: \mathbf{P}^2 = 2E$, and system trajectories are represented by fixed points on the sphere. Consider now the regime of  weak perturbation,  $\varepsilon \ll 1$. This implies that the system phase space $\mathbf{\Omega}$ remains almost completely filled with invariant tori, which form the set $\mathbf{\Omega}_{\text{tori}}$. The motion is regular on  the manifold $\mathbf{\Omega}_{\text{tori}}$ but there also appears a tiny manifold, the relative complement of $\mathbf{\Omega}_{\text{tori}}$ in $\mathbf{\Omega}$, which constitutes the \textit{Arnold web}, $\mathbf{\Omega}_{\text{web}} = \mathbf{\Omega} \backslash \mathbf{\Omega}_{\text{tori}}$. The Arnold web covers the resonance lines $\Omega_\mathbf{m}:=\{\mathbf{P}|\mathbf{m}\cdot\mathbf{P} = 0\}$,  where $\mathbf{m} = (m_x, m_y, m_z)$ is a triplet of coprimes, by  narrow chaotic channels. A typical channel width depends on the order of the corresponding resonance, $m = \max|m_\alpha|$, and it usually  decreases with the increase of $m$. The total volume occupied by the Arnold web is expected to scale as $\sqrt{\varepsilon}$ \cite{web}.

The appearance of the Arnold diffusion may take a while. The motion along the  web can  be detected only when the change $\triangle P(t) = \| \mathbf{P}(t) - \mathbf{P}(0)\|$ assumes a noticeable value. The Nekhoroshev theorem predicts that such a change can be observed after a time $t_{A}$, which scales at least exponentially with $1/\varepsilon$ \cite{neh}. Moreover, only a tiny fraction of the initial conditions, which entered into the manifold $\mathbf{\Omega}_{\text{web}}$, can diffuse. Therefore, even the numerical detection of the Arnold diffusion is a rather difficult and time-consuming task\cite{numericA, numericA1}.

Here, we consider a perturbation Hamiltonian of the explicit form:
\begin{equation}
H_p(x,y,z)= \cos(x)\cos(y)[1 + \cos(2 z)]. \label{eq2}
\end{equation}
The total Hamiltonian $H(\mathbf{P},\mathbf{X})$ in eqs. (\ref{eq1}, \ref{eq2})  cannot be separated into several independent two- or one-dimensional Hamiltonians; therefore the system is manifestly three-dimensional. The system in eqs. (1 - 2) falls within the range of Nekhoroshev theorem since the the Hamiltonian $H_0$ is convex and the function $H_p(\mathbf{X})$ is analytic. Moreover, the perturbation potential (\ref{eq2}) is bounded, $|H_p(\mathbf{X})| \leq 2$, and the total system energy, $E = H(\mathbf{P},\mathbf{X})$, effectively controls the strength of perturbation.
Therefore we set $\varepsilon=1$ and introduce an effective perturbation parameter
\begin{equation}
\varepsilon_{\text{eff}} = \max|H_p(\mathbf{X})|/H(\mathbf{P},\mathbf{X}) = 2/E\;. \label{eq3}
\end{equation}

In the momentum subspace the system evolves within the sphere $\mathbf{S}$. Namely, the system dynamics is confined to a thin layer enclosing the surface of the sphere $\mathbf{S}$, $L_E(\mathbf{P})$: $2(E - \varepsilon_{\text{eff}}) \leq \mathbf{P}^2 \leq  2(E + \varepsilon_{\text{eff}})$. Resonance lines $\mathbf{m}\cdot\mathbf{P} = 0$ form a set of circles on the surface of the sphere. In the limit $\varepsilon_{\text{eff}} \ll 1$, the Arnold web represents a sparse net on $\mathbf{S}$. Upon increasing the perturbation strength, $\varepsilon_{\text{eff}}$, the Arnold web tends to become `thicker': the invariant tori, which were located far outside of the low-order resonances (the distance is defined by the Diophantine resonance conditions \cite{web}), and were unaffected by the weak perturbation, become now modified.  The increasing perturbation involves more and more resonances into the growing Arnold web. This all results in the appearance of a complex, fine-structured network braiding  whole areas on the  surface of the sphere.

\section{Computational method}
Numerical studies of Arnold diffusion demand long runs. Therefore, one should resort to the integrators which consistently respect the symplectic properties of the original  systems \cite{symp}. A symplectic numerical scheme replaces the original, continuous-time Hamiltonian by its discretized version, and the numerical propagation produces the exact time evolution  of the corresponding Hamiltonian map \cite{foot1}. The so discretized  system dynamics thus is  a slightly perturbed version of the original one, so that there is no secular growth of the system integrals of motions, which is the total system energy $E$ in our case. The smaller the time step of the corresponding Hamiltonian map is, the closer both systems, original and discretized version stay close to each other. For the propagation of the Hamiltonian in eqs. (1 - 2)  we used the sixth-order symplectic integrator from Ref. \cite{symp_a}. For the time step $h = 10^{-2}T$, where the  timescale is determined by the characteristic frequency of the particle oscillations at the bottom of the potential well, $T = 2\pi$, the absolute error in the system energy, $\triangle E(t) = |E(t) - E(0)|$, did remain below $10^{-8}$ during the whole integration time.

All our calculations have been performed on a NVIDIA TESLA M2050 GPU, with 448 processing units on board, and  by using CUDA and double-precision accuracy. We employed the complete parallelization,  which allowed us in total to run $N = 10^7$ realizations simultaneously.  The performance gain is shown in Figure 1. While the calculation time on a single stack-structured CPU  grows linearly with $N$. At the same time, up to $N = 7168$ realizations  can be calculated on a GPU  simultaneously, due to the distribution procedure performed by CUDA. This number is a multiple of the number of GPU-cores, $7168 = 16 \times 448$. A further increase of the number of realizations causes a re-distribution of threads between the fixed number of cores and slow down the calculations speed, note the steps on the corresponding curve in Figure 1.

\begin{figure}[t]
\center
\includegraphics[width=0.48\textwidth]{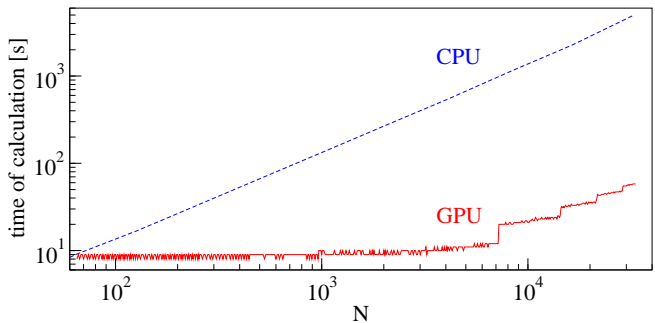}
\caption {Performance of a central processing unit (CPU) of the Augsburg University computational cluster (Intel Xeon Processor 2.93GHz Quad-Core X5570 Gainestown) versus the  GPU (NVIDIA TESLA M2050) for the propagation of an ensemble of Hamiltonian systems (\ref{eq1}, \ref{eq2}): the overall calculation time is depicted as a function of the number of realizations $N$ (see text for more details). The propagation time of a single realization was $t = 5000 T$.} \label{fig1}
\end{figure}

\begin{figure}[t]
\center
\includegraphics[width=0.48\textwidth]{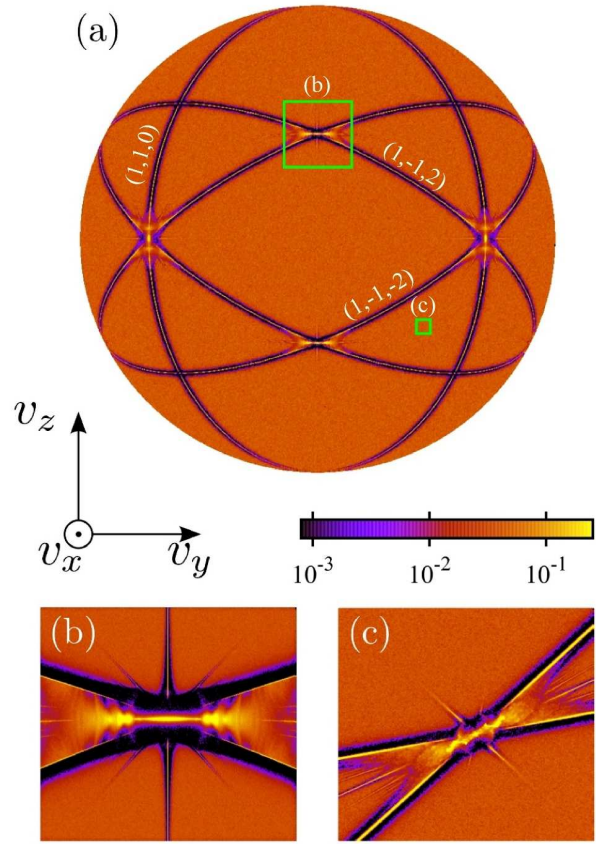}
\caption {(a) \textbf{Arnold web.} Probability density, $F_{\Delta t}(\mathbf{V};0)$ (see text), of the average velocity over $\Delta t =50T$ of the system in eqs. (1-2) starting from uniformly distributed initial conditions with a total energy $E = 400$. The braided stripes correspond to the resonances $\mathbf{m}\cdot\mathbf{P} = 0$.  The width of a stripe depends on the order of the resonance: the higher  the order, the narrower is the stripe.  Insets (b) and (c): These depict zooms of the corresponding resonance intersections. The inner stripe regions correspond to chaotic channels of the Arnold web, see the main text for more details.} \label{fig2}
\end{figure}

\section{Main  results of the simulations}
In order to visualize the Arnold web we employed the following procedure.  After having the initial conditions distributed over a certain region of the system phase space the ensemble of trajectories was launched.  The location and the shape of the distribution are tunable, so that we can scan different regions of the phase space at will. For every  trajectory from the ensemble, $\{ \mathbf{X}^i(t) \}$, $i = 1,..,N$, we  calculate the vector of the averaged finite-time velocity, $\mathbf{V}^i(t)=(V^i_x(t),V^i_y(t),V^i_z(t))$, with the components reading: $V^i_\alpha(t) = [X^i_\alpha(t+\Delta t) - X^i_\alpha(t)]/\Delta t$, wherein the averaging interval is given by $\Delta t$. In other words, it corresponds to a point on  $ \mathbf{S}$\/ \cite{foot2}. By running many such realizations in parallel, we collected the statistics,  and finally projected a probability distribution function, $F_{\Delta t}(\mathbf{V};t)$, on the surface of  the sphere  $\mathbf{S}$. -- The distribution is a nonstationary function, in the sense that its shape depends on $t$, so that the distribution profile reflects the dynamics of the ensemble.

If one of the ensemble trajectories  was launched from a region filled  with regular trajectories, it remains at the close vicinity of the initial momentum vector. There is no mixing in regular regions. Therefore the projection of the corresponding part of the initial probability density distribution onto the momentum sphere remains invariant.  When one of the ensemble trajectories entered the thin chaotic layer along a resonance $\mathbf{m}$, it stays within the channel for some time (``sticking event''), and during this time the system moves with the velocity vector, $\widetilde{\mathbf{P}}$, which nearly exactly obeys the resonance condition, $\mathbf{m}\cdot \widetilde{\mathbf{P}} \approx  0$. All the realizations within the chaotic layer contribute to the probability distribution, $F_{\Delta t}(\mathbf{V};t)$,  concentrated on the corresponding resonance. The resonance appears as  a bright line enclosed by an empty `dark zone', and the width of the dark zone  corresponds to the width of the web around the resonance. An increase of the averaging time  will induce further localization of the distribution at the corresponding resonance lines, but the widths of  channels, i. e. dark zones on the momentum sphere, are fixed and do not depend on $\Delta t$. 
Therefore, the color representation of $F_{\Delta t}(\mathbf{V};t)$ allows one to clearly identify the location of the Arnold channels.

 At this point, it is worth to refer to the two-dimensional limit of the Hamiltonian (\ref{eq1} - \ref{eq2}).
The momentum sphere represents a circle in this limit, and alternating chaotic and regular zones provide a full partition of the  circle into a set of sectors. A trajectory, being placed initially into one of the zones, stays  forever within the corresponding sector. Each chaotic zone can be characterized by the corresponding asymptotic velocity, $\mathbf{V^i}=(V^i_x,V^i_y)$. When the averaging time $\Delta t$ approaches infinity, the distribution function inside the corresponding chaotic sector shrinks to the point $\mathbf{V^i}$. Therefore, in the asymptotic limit  sectors, corresponding to chaotic zones look like dark regions with  bright points at their centers.     

The averaging time interval $\Delta t$ is tunable and there exists no {\it a priori} best choice for it. Namely, every resonance channel $\mathbf{m}$ can be characterized by a probability distribution of the corresponding sticking times, $\psi_{\mathbf{m}}(t_{\text{stick}})$. The distribution always possesses a finite first moment\cite{kac}, a mean sticking time $t_{\text{stick}}(\mathbf{m}) = \int_0^\infty \tau \psi_{\mathbf{m}}(\tau) d\tau$. If the averaging time interval is much larger than this mean sticking time,  the corresponding resonance channel cannot be resolved. If, on the  contrary,  $\Delta t \ll t_{\text{stick}}(\mathbf{m})$, the oscillations of the momentum vector $\mathbf{P}(\mathbf{m})$ along the high-order resonance, with $m\gg 1$, will smear the probability density over a broad region thus preventing  a resolution of the fine structure of the web. In the following  we used the value $\Delta t=50T$.

The high degree of parallelism  capability of the GPU offers a massive speedup  for the ensemble propagation, see in Figure~1. With our  Figures~2 and 3 we show the results  of our numerical experiment with $N \approx 10^8$ independent realizations and the propagation time has been set at $t_p = \Delta t$. The overall simulation time of each experiment assumed 45 minutes. For a very long time evolution, see in Figure~4,  we propagated an ensemble of $N = 3.2 \cdot 10^4$  particles up to a time $t_p = 10^7T$. The simulation time in this case was 24 hours. The standard CPU-based run of a numerical experiment of the same scale would require (i) approximately \textbf{one year} of calculations on  a  standard desktop PC or (ii) about  \textbf{five months} on a  more advanced CPU of the Augsburg University  computational cluster.

The results corresponding to the weak perturbative regime  at a total energy $E = 400$, corresponding to $\varepsilon_{\text{eff}}=0.005$, are shown in Figure~2. We used an ensemble with the initial conditions uniformly distributed over the sphere  $\mathbf{S}$, and over the torus $[0, 2\pi]\times [0, 2\pi] \times [0, 2\pi] $ in the coordinate space $\mathbf{\Omega}_{\mathbf{X}}$. The most of the sphere surface is occupied by  regular trajectories, therefore the initial uniform distribution remains invariant almost everywhere, representing the uniform dark orange area on Figure 2a.  The brightest narrow stripes which braid the sphere correspond to the lower order, $m \leq 2$, resonances. We also scanned the energy shell by using an ensemble of trajectories with the initial conditions uniformly distributed within a small segment of the sphere (enclosed by the two squares in Fig.~2~(a), with the aim to resolve the presence of the Arnold web around some higher-order resonances. The results, depicted in Figs.~2~(b,c),  show that the structures of the resonance intersections are topologically similar.

\begin{figure}[b]
\center
\includegraphics[width=0.48\textwidth]{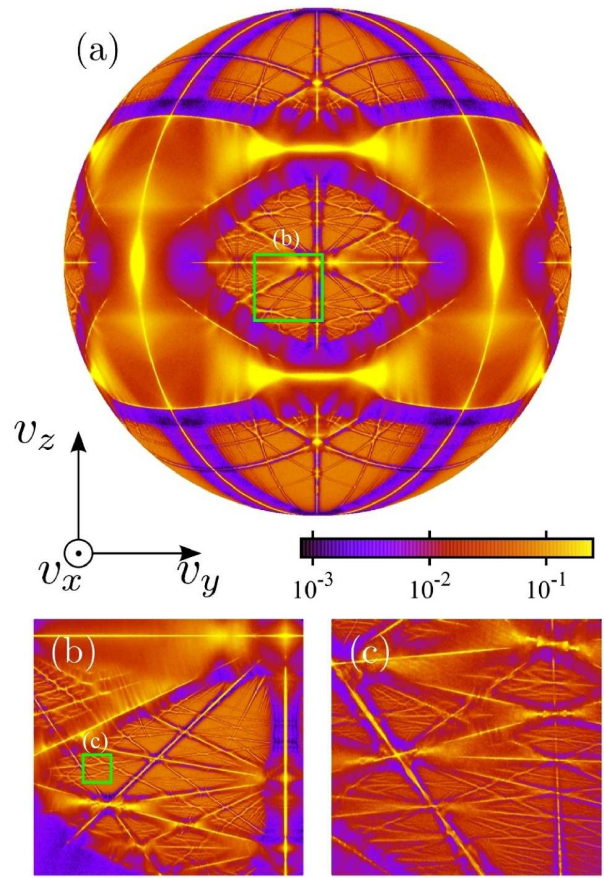}
\caption {\textbf{Arnold web.} Probability density, $F_{\Delta t}(\mathbf{V};0)$ (see text), of the average velocity over $\Delta t =50T$ of the system in eqs. (1-2) starting from uniformly distributed initial conditions with a total energy $E = 15$. The phase space is pierced by the resonance channels of  different orders, which form fine structured patches on the velocity sphere, see the sequel of  insets from (a)
$\rightarrow $ (b) $\rightarrow$ (c). The bright areas correspond to the zones of well-developed chaos. For the highest resolution in panel (c) we double the averaging interval $\Delta t=100T$.} \label{fig3}
\end{figure}

\begin{figure*}[t]
\leavevmode\includegraphics[width=0.95\hsize]{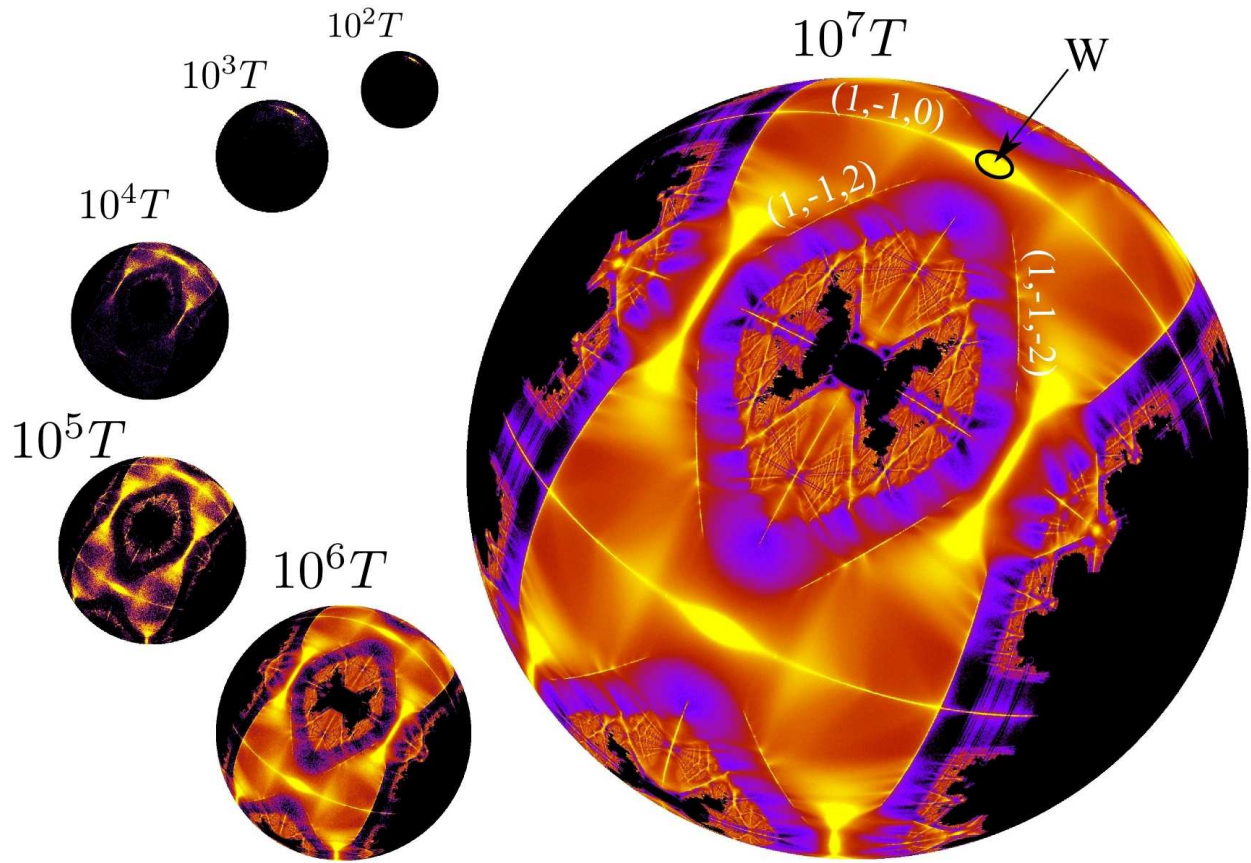}
\caption{\textbf{Arnold diffusion.} The time evolution of the probability density, $F_{\Delta t}(\mathbf{V};t)$, depicts the spreading of the ensemble of $N = 3.2 \cdot 10^4$ trajectories localized initially at the intersection of three major resonances,  $\mathbf{W} = \Omega_{(1,-1,2)} \cap \Omega_{(1,-1,-2)} \cap \Omega_{(1,-1,0)}$. The initial cloud assumes a Gaussian  profile in the velocity space, with the center at the point $\mathbf{W}$ and dispersions $\sigma_{v_x} = \sigma_{v_y} = 0.01$, and a uniform distribution over the torus $[0, 2\pi]\times [0, 2\pi] \times [0, 2\pi] $ in the coordinate space, $\mathbf{\Omega}_{\mathbf{X}}$. The corresponding initial velocity $v_z$ was calculated from the fixed energy condition $E=15$.  To resolve the fine structure of the web with a relatively small ensemble we averaged the probability density $F$ over the whole time span between  consecutive plots. Altough the radius of all spheres in the momentum space is the same, $\sqrt{2E}$, we  consequently increased their sizes in order to resolve the fine structure of the developing Arnold web.}
\label{fig4}
\end{figure*}

The results for the regime of moderate perturbation with total energy $E = 15$; i.e.  $\varepsilon_{\text{eff}}=2/15$ , are depicted with Fig.~3. Although the perturbation  is much stronger now, it is not yet strong enough so as to destroy all the resonance tori and bring the system into the regime of the global Chirikov diffusion \cite{chir}. There are several patches of the resonance web, which  form  leaf-like clusters on the  surface of the sphere.  The zooms into clusters reveal a fine fractal structure, see Fig.~3~(b,c). The clusters are formed by higher-order resonances, and the width of the corresponding channels scales  with the increase of the resonance orders. Therefore, both the number of realizations, $N(m)$, and the averaging time, $\Delta t(m)$, needed to resolve the web structure formed around the resonances of order $m$,  grow quickly with $m$. In order to overcome this obstacle we scan the region of interest with the ensemble launched within the designated area, and then tracked only those trajectories which remained within the area during the whole observation time. The results are shown in Figures~3~(b,c).

Finally, we computed  the diffusive spreading of the ensemble initially injected at the point of intersection of three  low-order resonances, $\mathbf{W} = \Omega_{(1,-1,2)} \cap \Omega_{(1,-1,-2)} \cap \Omega_{(1,-1,0)}= (\sqrt{E}, \sqrt{E},0)$, which is at the center of a well-developed chaotic region. The evolution dynamics proceeds in a stepwise manner. In a first stage we detect a  fast spreading over the chaotic region, which encloses the injection point. Then the spreading slows down, and eventually enters a second stage of slow diffusion through the Arnold web. Both stages are illustrated on Figure~4 \cite{foot3}. The system phase space is not uniform so that two dynamical stages correspond to two different regions of the energy shell. The region at the  vicinity of the intersections of the primary resonances is strongly chaotic, and the dynamics there is governed by the fast Chirikov diffusion. The rest of the shell is well-structured by the KAM tori, and the evolution there is mediated by the Arnold web. Fractal clusters play the role of hubs between two regions.

\section{Conclusions} With this numerical study we have shown how a GPU  supercomputer can be used to explore the Arnold diffusion in near-integrable Hamiltonian systems.  The appearance  of the Arnold diffusion  demands the evaluation of  huge statistical data sets. It may be considered as a typical problem of sampling of rare events \cite{numericB}. Namely, once the system got into a narrow resonance channel, it stayed there for a while before making a transition to another channel. These transitions constitute  rare events, which in fact  determine the long-time evolution of the system.  The disparity of the involved time scales, which strongly depend on the resonance orders, makes the mapping of the diffusion web a very challenging computational task.  The use of the CUDA-based NVIDIA platform led to speedups by the  factors $\sim 350$ and $\sim 100$, as compared  with the performances of a standard PC  and of a computational cluster's CPU, correspondingly.

When compared to other, more sophisticated methods \cite{numericA, numericAa, numericA1, numericB}, our approach to the visualization of the Arnold diffusion in three-dimensional autonomous Hamiltonian systems revails some advantages, both in theoretical and experimental domains. First, our scheme  neither demands pre- nor after-processing but only a straightforward integration of the corresponding equations of motion. Second, it allows for a direct extension to the quantum limit  while it is not at all clear how one could generalize the concept of finite-time Lyapunov exponents \cite{numericA, numericA1, numericB} or the frequency analysis \cite{numericAa} to the Schroedinger equation.

The state-of-the-art experiments with ultracold atoms provide a natural playground for the realization of our method. The creation of three-dimensional periodic optical potentials nowadays become an experimental routine \cite{ober}. The initial ensemble in a form of  narrow distribution over a certain manifold in the three-dimensional  momentum space can be prepared by using a diluted BEC cloud and the Bragg selection technique \cite{bregg1, bregg2}. The needed  time, $\Delta t \lesssim 100T$, where $T$ is the period of oscillations at the bottom of potential well, is several orders of magnitude smaller then the characteristic decoherence time \cite{ober}. Therefore, the  evolution of ultracold atoms is  Hamiltonian. The  instantaneous momentum distribution can be measured  by using the standard time-of-flight measurement technique \cite{ober}. Finally, a tunable effective Planck constant allows to probe both the semiclassical and the deep quantum limits. 

Another intriguing  application that comes to mind  is the implementation of the GPU-based mapping procedure to search for the footprints of the Arnold diffusion in the emission patterns of three-dimensional optical resonators \cite{resonator}.

Our work here illustrates the advantages, provided by GPC -- GPU ideology, for nonlinear dynamics studies by using a particular example. Yet the potential of this approach  reaches far beyond: it has already been put to work to explore (i) the noisy phase dynamics in a Josephson junction and the noisy Kuramoto model \cite{gpu4}, (ii) the long-time evolution of  nonlinear lattices \cite{zab},  and (iii) the functioning of inertial Brownian motors that are driven by colored noise \cite{motor}.  We thus believe that the use of  GPU computing in nonlinear science is only at a beginning: in the immediate future its great potential likely will provide many unforeseen findings and possibly even unravel manifest new phenomena.

\section{acknowledgements}
This work has been supported by the DFG Grant HA1517/31-2 (S.D. and P.H.).


\begin{thebibliography}{10}
\bibitem{foot1} Numerical evolution is `exact' up to the round-off error, which is $10^{-32}$ for double-precision calculations.
\bibitem{foot2} More precisely, it corresponds to a point \textit{within} the thin layer beneath the sphere surface, $L_E(\mathbf{P})$. However, the layer width is small for the range of the energies we will consider here, so finally we use the central projection of the point $\mathbf{V}$ on the  surface of the sphere.
\end{thebibliography}
\end{document}